\documentclass[twocolumn,aps,prl]{revtex4}
\usepackage{graphicx}

\begin{document}
\bibliographystyle{prsty}

\title{Time-Resolved Quasiparticle Dynamics in the Spin-Density-Wave State}
\author{Elbert E. M. Chia}
\author{Jian-Xin Zhu}
\author{H. J. Lee}
\altaffiliation{Department of Physics, University of California at
Berkeley, California, USA}
\author{Namjung Hur}
\altaffiliation{Department of Physics, Inha University, Incheon
402-751, South Korea}
\author{N. O. Moreno}
\altaffiliation{Department of Physics, Federal University of
Sergipe, S$\tilde{a}$o Cristov$\tilde{a}$o, SE 49100-000, Brazil}
\author{R. D. Averitt}
\author{J. L. Sarrao}
\author{A. J. Taylor}
\affiliation{Los Alamos National Laboratory, Los Alamos NM 87545,
USA}
\date{\today}

\begin{abstract}
Time-resolved photoinduced reflectivity is measured in the
spin-density-wave (SDW) phase using itinerant antiferromagnets
UMGa$_{5}$ (M=Ni, Pt). For UNiGa$_{5}$ [$T_{N}$=85~K,
$Q$=($\pi$,$\pi$,$\pi$)], the relaxation time $\tau$ shows a sharp
increase at $T_{N}$ consistent with the opening of a SDW gap. For
UPtGa$_{5}$ [$T_{N}$=26~K, $Q$=(0,0,$\pi$)], no change in $\tau$ is
observed at $T_{N}$ or at the lowest temperatures. We attribute this
to the absence of the SDW gap at the Fermi level, due to a different
modulation vector $Q$, which leads to a gapless quasiparticle
spectrum. Our results challenge the conventional wisdom that a SDW
phase necessarily implies a SDW gap at the Fermi level.
\end{abstract}

\maketitle

The quasiparticle (QP) dynamics of the spin-density-wave (SDW) phase
is an important area of study, especially in the cuprate
high-temperature superconductors (HTSC), where superconductivity and
antiferromagnetism (AFM) could coexist, either in an applied
magnetic field \cite{Lake02,Kang03}, or in zero field
\cite{Mukuda06}. In single-layered non-superconducting cuprates,
where there is only a single layer of CuO$_{2}$ plane in the unit
cell, AFM is necessarily G-type, where the nearest-neighbor spins in
the CuO$_{2}$ plane are antiferromagnetically (AF)-aligned. However,
in multi-layered cuprates, two types of AFM can occur on the
CuO$_{2}$ planes: (1) G-type AFM, or (2) A-type AFM, i.e. spins are
ferromagetically (FM) aligned along each plane, but spins on
adjacent planes are AF-aligned. Being able to elucidate the magnetic
alignment of the spins in multi-layered cuprate HTSCs in the
coexistence phase is crucial in narrowing down the starting point
for a pairing theory in the HTSCs. It is therefore important to
study the QP dynamics of pure SDW systems before proceeding to more
complex coexistence phases in HTSCs. At present there have been no
systematic measurements of the QP dynamics of the SDW phase,
\textit{as a function of the type of AFM}, where below $T_{N}$, a
SDW gap, i.e. a charge gap on the nested parts of the Fermi surface
(FS), might open up.

Usually, the SDW state in itinerant AFMs is probed by techniques
such as resistivity, specific heat and neutron scattering. In
resistivity and specific heat, a feature appearing at $T_{N}$, such
as a hump or peak, has been interpreted as due to the formation of a
SDW gap $\Delta_{SDW}$ and the accompanying partial disappearance of
the FS (See, for example,
Ref.~\onlinecite{Tokiwa01,Tokiwa02a,Fawcett88}). However, it is not
clear if the feature at $T_{N}$ is due to an \textit{actual} gap
opening up in the DOS at the Fermi level [DOS($E_{F}$)], or is
merely due to a decrease in DOS($E_{F}$) without it vanishing. In
elastic neutron scattering (ENS), the intensity of the neutron
scattering peak $I_{NS}$ increases from zero below $T_{N}$ in a
BCS-like manner, and is proportional to $M^{2}$, where $M$ is the
staggered magnetization. In an itinerant AFM, Overhauser
\cite{Overhauser62} derived $M \propto \Delta_{SDW}$, yielding
$\Delta_{SDW} \propto \sqrt{I_{NS}}$. However, does the presence of
a staggered magnetization \textit{always} imply the presence of the
gap in the DOS? Looking at this problem from another perspective,
does the opening up of a gap depend on the \textit{type} of AFM in
the material?

Ultrafast optical spectroscopy (UOS) has recently been used in the
study of correlated electron materials. For example, in heavy
fermions (HF) such as YbAgCu$_{4}$ \cite{Demsar03a} and CeCoIn$_{5}$
\cite{Demsar06b}, time-resolved photoinduced reflectivity
measurements display a divergence of the electron-phonon (e-ph)
relaxation time $\tau$ at the lowest temperatures $T$. In materials
with a gap in the QP spectrum such as HTSCs like YBCO
\cite{Han90,Demsar99c} and charge-density-wave materials like
K$_{0.3}$MoO$_{3}$ \cite{Demsar99b}, $\tau$ diverges near $T_{c}$
when a gap opens in the QP density of states (DOS). The $T$
dependence of the relaxation time and peak amplitude has been
explained by the phenomenological Rothwarf-Taylor (RT) model. This
model describes the relaxation of photoexcited SCs \cite{Demsar03c},
where the presence of a gap in the electronic DOS gives rise to a
relaxation bottleneck for carrier relaxation, arising in SCs from
the competition between QP recombination and pair breaking by
phonons \cite{Demsar01}.

In this Letter, we investigate the QP response in the SDW state
using UMGa$_{5}$ (M=Ni, Pt) as a model system. For UNiGa$_{5}$, the
decay time of $\Delta R/R$ (directly related to $\tau$) increases
sharply at $T_{N}$ and shows a quasi-divergence below $T_{N}$,
consistent with the opening of a SDW gap. For UPtGa$_{5}$, however,
there is no discernible change in $\tau$ across $T_{N}$ and at the
lowest $T$. We attribute this to the QP spectrum being gapless,
which is the result of the SDW modulation vector $Q$ being different
from that of UNiGa$_{5}$. Our technique thus enables us to
distinguish the SDW state of two very similar materials, one with a
SDW gap (UNiGa$_{5}$), and the other without a SDW gap
(UPtGa$_{5}$). We further substantiate our claim by performing a
microscopic model calculation on a 3-dimensional (3D) cubic lattice
in the SDW state, and show that the presence or absence of a SDW gap
is intimately related to the value of $Q$ of that particular
material. Our results challenge the conventional wisdom that a SDW
phase necessarily implies a SDW gap at the Fermi level.

UMGa$_{5}$ (M=Ni, Pt) are 5\textit{f} itinerant AFMs with
$T_{N}$$\approx$85~K (Ni) and 26~K (Pt), respectively. The AF phase
in UNiGa$_{5}$ is G-type, i.e. the nearest-neighbor spins are
AF-aligned, with modulation vector $Q$=($\pi$,$\pi$,$\pi$)
\cite{Tokiwa02c}. Contrast this with UPtGa$_{5}$, where the AF phase
is A-type, i.e. the spins are FM-aligned in the \textit{ab} plane
and AF along the \textit{c} axis, with $Q$=(0,0,$\pi$)
\cite{Tokiwa02c}. The electronic specific heat coefficients $\gamma$
= 30~mJ/mol.K$^{2}$ (Ni), and 57~mJ/mol.K$^{2}$ (Pt), respectively
\cite{Tokiwa01,Tokiwa02a}. The moderate values of $\gamma$ suggest
that these materials are marginal HFs.

Single crystals of UMGa$_{5}$ were grown in Ga flux \cite{Moreno05},
with dimensions $\sim$1 x 1 x 0.4mm$^{3}$. Specific heat
measurements were performed in a Quantum Design PPMS from 2~K to
300~K to determine $T_{N}$. The shape of the specific heat anomaly
at $T_{N}$ is similar for both materials. The photoinduced
reflectivitiy measurements were performed using a standard
pump-probe technique \cite{Demsar99d}, with a Ti:sapphire laser
producing sub-100~fs pulses at approximately 800~nm (1.5~eV) as the
source of both pump and probe optical pulses. The pump and probe
pulses were cross-polarized. The experiments were performed with a
pump fluence of $<$1.0 $\mu$J/cm$^{2}$, yielding a photoexcited QP
density $n_{pe}$$<$0.05/unit cell. The probe intensity was $\sim$25
times lower. Data were taken from 10~K to 300~K. The photoinduced
$T$ rise at the lowest $T$ was estimated to be $\sim$10~K for
UNiGa$_{5}$ and $\sim$6~K for UPtGa$_{5}$ (accounted for in all data
plots).

\begin{figure} \centering \includegraphics[width=8cm,clip]{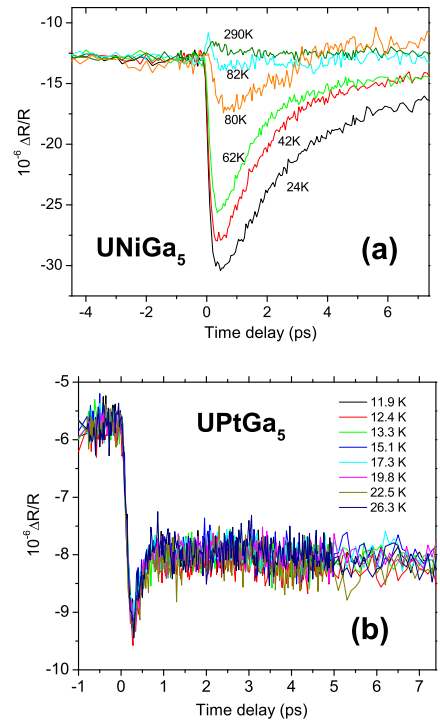}
\caption{Transient reflection $\Delta R/R$ after photoexcitation by
a 100-fs laser pulse above and below $T_{N}$, for (a) UNiGa$_{5}$,
(b) UPtGa$_{5}$.} \label{fig:NiPtFig1}
\end{figure}

In Figure~\ref{fig:NiPtFig1} we show the time dependence of the
photoinduced signal of UNiGa$_{5}$ and UPtGa$_{5}$ below and above
$T_{N}$. The time evolution of the photoinduced reflection $\Delta
R/R$ first shows a rapid rise time (of the order of the pump pulse
duration) followed by a subsequent picosecond decay. These data can
be fit using a single exponential decay over the entire $T$ range,
$\Delta R/R = A \exp (-t/\tau)$.

\begin{figure}
\centering \includegraphics[width=8cm,clip]{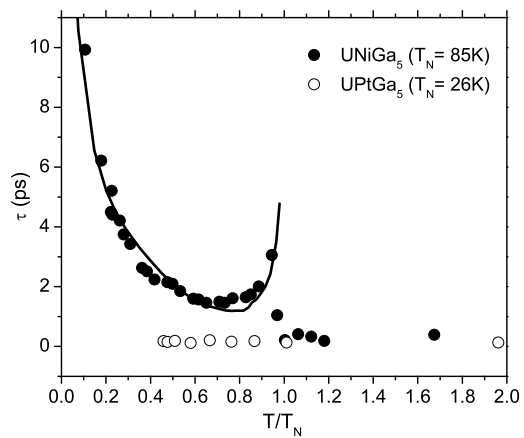} \caption{$T$
dependence of relaxation time $\tau$ for (solid circles)
UNiGa$_{5}$, (o) UPtGa$_{5}$. Solid line: Fit of UNiGa$_{5}$ data
using Rothwarf-Taylor model, taken from Ref.~\onlinecite{Chia06}.}
\label{fig:NiPtFig2}
\end{figure}

Figure~\ref{fig:NiPtFig2} shows the $T$ dependence of the relaxation
time $\tau$ for UNiGa$_{5}$ and UPtGa$_{5}$ extracted from
Fig.~\ref{fig:NiPtFig1}. For UNiGa$_{5}$ (solid circles) $\tau$
shows an abrupt increase near $T_{N}$ followed by a gradual increase
at the lowest $T$. The QP dynamics of $\tau$ has been explained by
us in Ref.~\onlinecite{Chia06} using the RT model, where the opening
of the SDW gap, which is the charge gap that opens up along the
nested regions of the Fermi surface (FS), leads to a relaxation
bottleneck. In contrast, for UPtGa$_{5}$ (open circles), there is no
discernible change in $\tau$ across $T_{N}$ or at the lowest $T$,
implying that no SDW gap opens up in the AF phase. In fact, this $T$
dependence is similar to what has been observed in conventional
wide-band metals \cite{Groeneveld95}.

We propose a theoretical model below to explain why a SDW gap opens
up in UNiGa$_{5}$ but not in UPtGa$_{5}$. We attribute the absence
of a SDW gap to be a direct consequence of the value of the AF
modulation vector $\bf{Q}$=(0,0,$\pi$) in UPtGa$_{5}$, compared with
($\pi$,$\pi$,$\pi$) in UNiGa$_{5}$. The model Hamiltonian for the
SDW can be expressed as
\begin{equation}
\mathcal{H} = \sum\limits_{\bf{k},\sigma} \epsilon_{\bf{k}}
c^{\dag}_{\bf{k} \sigma} c_{\bf{k} \sigma} + \frac{U}{N}
\sum\limits_{\bf{k},\bf{k^{\prime}}} c^{\dag}_{\bf{k} \uparrow}
c_{\bf{k^{\prime}} \uparrow} c^{\dag}_{\bf{k^{\prime}+Q} \downarrow}
c_{\bf{k+Q} \downarrow} \label{eqn:Hamiltonian}
\end{equation} where $c^{\dag}_{\bf{k} \sigma} (c_{\bf{k} \sigma})$
is the creation (annhilation) operator of an electron having the
wavenumber $\bf{k}$ and spin $\sigma$. The first term represents the
one-electron energy with dispersion $\epsilon_{\bf{k}}$, which, for
a 3D cubic lattice, is given by
\begin{equation}
\epsilon_{\bf{k}} = -2t (\cos k_{x}a + \cos k_{y}a + \cos k_{z}a)
\label{eqn:epsilonk}
\end{equation} where $t$ is the overlap integral between the
5\textit{f} wavefunctions on neighboring sites, and $a$ is the
lattice spacing. The second term in $\mathcal{H}$ denotes the
on-site Coulomb interaction, with $U$ being the on-site Coulomb
energy, and $N$ the number of lattice sites.

\begin{figure}
\centering \includegraphics[width=8cm,clip]{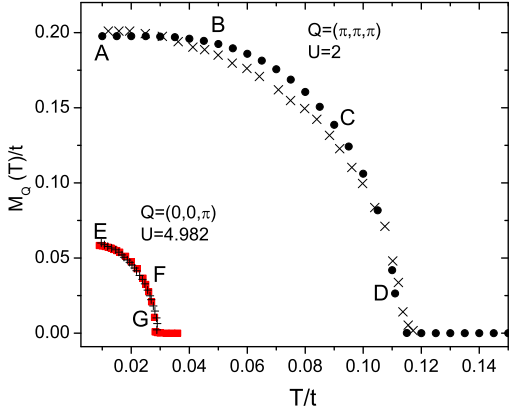}
\caption{Calculated $T$-dependence of the order parameter $M_{Q}(T)$
for a 3D cubic lattice for $Q$=($\pi$,$\pi$,$\pi$) (solid circles),
and $Q$=(0,0,$\pi$) (solid squares). Experimental ENS intensity for
(+) UPtGa$_{5}$, and ($\times$) UNiGa$_{5}$, taken from
Ref.~\onlinecite{Tokiwa02c}.} \label{fig:MQ}
\end{figure}

We define the SDW order parameter $M_{Q}$
\begin{equation}
M_{Q} \equiv \frac{-U}{N} \sum\limits_{\bf{k}} \langle
c^{\dag}_{\bf{k+Q} \downarrow} c_{\bf{k} \uparrow} \rangle.
\label{eqn:SQ}
\end{equation} Then, after performing unitary transformation, bilinearization and
diagonalization, we obtain the $T$ dependence of $M_{Q}$ in the
following self-consistent equation
\begin{equation}
M_{Q} = \frac{U}{N} \sum\limits_{\mathbf{k} \in rBZ}
\left\{\frac{-M_{Q}\left[f(E_{+}) -
f(E_{-})\right]}{\sqrt{\frac{1}{4} (\epsilon_{\mathbf{k}} -
\epsilon_{\mathbf{k+Q}})^{2} + M_{Q}^{2}}}\right\} \label{eqn:MQT}
\end{equation} where $f$ is the Fermi function, the $\bf{k}$-points are
taken from the reduced Brillioun Zone (rBZ), and $E_{+}$ and $E_{-}$
are the two branches of the one-particle energy in the SDW state
given by
\begin{equation}
E_{\pm}(\mathbf{k}) = \frac{1}{2}(\epsilon_{\mathbf{k}} +
\epsilon_{\mathbf{k+Q}}) \pm \sqrt{\frac{1}{4}
(\epsilon_{\mathbf{k}} - \epsilon_{\mathbf{k+Q}})^{2} + M_{Q}^{2}}.
\label{eqn:Ek}
\end{equation} Fig.~\ref{fig:MQ} shows the $T$ dependence of
$M_{Q}$ for $Q$=($\pi$,$\pi$,$\pi$) (solid circles) and (0,0,$\pi$)
(solid squares). $U$ for each case is chosen such that the ratio
$T_{N}$[$Q$=(0,0,$\pi$)]/$T_{N}$[$Q$=($\pi$,$\pi$,$\pi$)] is
$\sim$26/85. We see that the order parameter $M_{Q}(T)$ follows the
BCS $T$ dependence, as expected. Furthermore, this agrees with ENS
results of UNiGa$_{5}$ and UPtGa$_{5}$ \cite{Tokiwa02c} (shown as
``+'' and ``$\times$'' in Fig.~\ref{fig:MQ}), where the ENS
intensity is a measure of the sublattice magnetization $M_{Q}$.

For UNiGa$_{5}$, where $Q$=($\pi$,$\pi$,$\pi$), we obtain
$\epsilon_{\bf{k}}$=$- \epsilon_{\bf{k}-\bf{Q}}$ from
Eq.~\ref{eqn:epsilonk}, and thus $E_{+}$ and $E_{-}$ are separated
by a well-defined energy gap in all of $\bf{k}$-space. Contrast this
with UPtGa$_{5}$, with $Q$=(0,0,$\pi$) --- here $\epsilon_{\bf{k}}
\neq - \epsilon_{\bf{k}-\bf{Q}}$, and thus $E_{+}$ and $E_{-}$ are
no longer separated by a gap in all of $\bf{k}$-space.

\begin{figure}
\centering \includegraphics[width=8cm,clip]{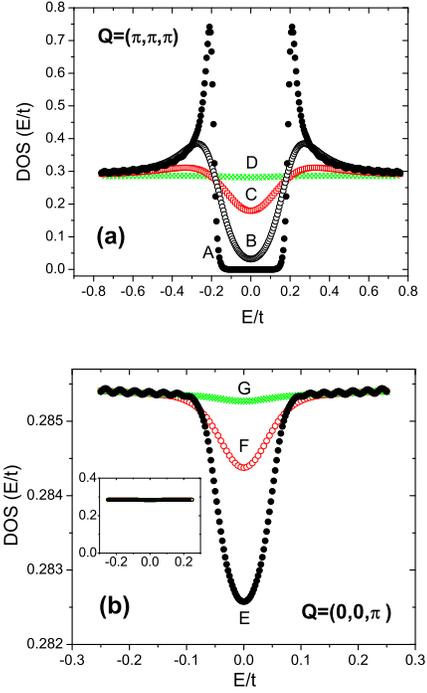}
\caption{Calculated $T$ dependence of the DOS in the SDW state for
(a) $Q$=($\pi$,$\pi$,$\pi$), and (b) $Q$=(0,0,$\pi$). The curves
A--G represents DOS calculated from the values of $T$ and $M_{Q}(T)$
in Fig.~\ref{fig:MQ}. $E$=0 is the Fermi level.} \label{fig:DOS}
\end{figure}

To further strengthen our case, we calculate the DOS and its $T$
dependence. It is given by
\begin{equation}
\rho (E) = \frac{-2}{N} \sum\limits_{\mathbf{k} \in rBZ}
\left\{f^{\prime}[E - E_{+}(\mathbf{k})] + f^{\prime} [(E -
E_{-}(\mathbf{k})]\right\} \label{eqn:DOS}
\end{equation} where $f^{\prime} \equiv \partial f(E)/\partial E$ accounts for
thermal smearing, and the factor 2 accounts for spin degeneracy.
Note that in Eq.~\ref{eqn:DOS}, $f^{\prime}$ becomes the
delta-function in the $T$=0 limit. The resulting DOS is shown in
Fig.~\ref{fig:DOS}, where the letters A--G represents the calculated
DOS based on the values of $T$ and $M_{Q}(T)$ in Fig.~\ref{fig:MQ}.
The Fermi level is located at $E$=0. For a $Q$=($\pi$,$\pi$,$\pi$)
SDW material, illustrated in Fig.~\ref{fig:DOS}a, a well-defined
energy gap develops below $T_{N}$, as evidenced by the sharp dips
and coherence peaks appearing at $\pm M_{Q}$. With decreasing $T$,
the dips approach zero, and the gaps are more well-defined due to
less thermal smearing. For $Q$=(0,0,$\pi$), however, though dips do
appear below $T_{N}$, the magnitude of the dips were very small
($\sim$10$^{-3}$), even near $T$=0. In the same scale as
Fig.~\ref{fig:DOS}a, it is almost flat, i.e. energy-independent, as
shown in the inset of Fig.~\ref{fig:DOS}b. We therefore conclude
that in UNiGa$_{5}$ [$Q$=($\pi$,$\pi$,$\pi$)], a well-defined energy
gap forms near the Fermi level in the SDW phase, whereas in
UPtGa$_{5}$ [$Q$=(0,0,$\pi$)], DOS is finite and almost constant
below and above the Fermi level. This therefore agrees with our
experimental results, where, in the SDW phase below $T_{N}$, we see
a charge gap opening up in UNiGa$_{5}$ but not in UPtGa$_{5}$.

Two pieces of experimental data might suggest the presence of a gap
in the QP spectrum in UPtGa$_{5}$, in contradiction to our results.
First, resistivity data $\rho (T)$ \cite{Tokiwa02a} showed a small
hump at $T_{N}$, similar to UNiGa$_{5}$ \cite{Tokiwa01}, which the
authors attributed to an opening of a gap in the band structure and
the partial disappearance of the FS. However, closer examination
reveals that for UPtGa$_{5}$, for both current directions [100] and
[001], $\rho$ did not \textit{increase} upon entering the AF phase,
unlike in UNiGa$_{5}$, where in the [001] direction, $\rho$
increases just below $T_{N}$, which is identical to the behavior of
Cr upon entering the AF phase. The hump seen in UPtGa$_{5}$ could be
merely due to a slight decrease in the QP DOS below $T_{N}$, as
shown in our analysis in Fig.~\ref{fig:DOS}b, without actually
forming a gap on the FS. We also emphasize that the low-$T$
($T<T_{N}$) $\rho (T)$ data in Ref.~\onlinecite{Moreno05} yields a
value of $\Delta$ that is \textit{not} the SDW gap, but is rather
the spin gap, i.e. a gap in the magnon dispersion \cite{Andersen79}.
Second, ENS intensity $I_{NS}$ \cite{Tokiwa02c} increases below
$T_{N}$ in a BCS-like manner, which may suggest that a SDW gap
$\Delta$ also opens up in the AF phase, since $\Delta \propto
\sqrt{I_{NS}}$ \cite{Overhauser62}. However, this expression
strictly only applies to G-type AF phase, where the nearest-neighbor
spins are AF-aligned as in UNiGa$_{5}$. We have shown in our
previous analysis that $\Delta \propto \sqrt{I_{NS}}$ is
\textit{not} true in the A-type AF state, where the spins are
FM-aligned in the \textit{ab} plane and AF-aligned along the
\textit{c} axis, as in UPtGa$_{5}$. $I_{NS}(T)$ from ENS merely
measures the order parameter in the SDW phase, i.e. the
\textit{staggered magnetization} $M_{Q}$. It strictly does not
measure the SDW gap. Hence our results do not contradict other
experimental results. According to the RT model, the lack of a gap
at the FS also explains why there is no upturn in $\tau$ at the
lowest $T$. Hence we have shown that due to a different modulation
vector $Q$ in UPtGa$_{5}$, a gap does \textit{not} open up at the
Fermi level, resulting in a lack of upturn in $\tau$ at both $T_{N}$
and at the lowest $T$.

It is interesting to note from Fig.~\ref{fig:NiPtFig2} that the
value of $\tau$ for UPtGa$_{5}$ is sub-ps, and is almost
$T$-independent. This is commonly seen in metals
\cite{Groeneveld95}, where the comparable electron-electron (e-e)
and e-ph relaxation rates results in the electron gas not being able
to reach thermal equilibrium long before e-ph energy relaxation
process sets in. An initial \textit{non-thermal} electron
distribution is thus a necessary starting point to derive the
$T$-independence and the sub-ps value of $\tau$. Since both
UNiGa$_{5}$ and UPtGa$_{5}$ are good metals, the value of their
$\tau$'s in the paramagnetic (PM) phase, the $T$-independence of
$\tau$ of UPtGa$_{5}$, as well as their similar values of DOS in the
PM phase (from Fig.~\ref{fig:DOS}), are consistent with a
non-thermal electron distribution immediately following
photoexcitation.

We have performed time-resolved photoinduced reflectivity
measurements in the SDW phase using itinerant AFMs UMGa$_{5}$ (M=Ni,
Pt) as model systems. For UNiGa$_{5}$ [$T_{N}$=85~K,
$Q$=($\pi$,$\pi$,$\pi$)], the relaxation time $\tau$ shows a sharp
increase at $T_{N}$ consistent with the opening of a SDW gap. For
UPtGa$_{5}$ [$T_{N}$=26~K, $Q$=(0,0,$\pi$)], no change in $\tau$ was
seen at both $T_{N}$ and at the lowest $T$. We attribute this to the
absence of the SDW gap at the Fermi level, due to a different
modulation vector $Q$, which leads to a gapless QP spectrum. Our
analysis also applies to 2D materials. Our study thus extends the
utility of UOS to study SDW materials, that enables us to
sensitively probe the presence or absence of a SDW gap in the AF
phase.

Work at Los Alamos was supported by the Los Alamos LDRD program.
E.E.M.C. acknowledges G. T. Seaborg Postdoctoral Fellowship support.

\bibliographystyle{prsty}
\bibliography{UMGa5,Ultrafast,PMGa5,CeCoIn5v11,RNBC,Pb,PrOs4Sb12}
\bigskip

\end{document}